# Rabi resonance splitting phenomena in photonic integrated circuits

*David J. Moss*

[1]Optical Sciences Centre, Swinburne University of Technology, Hawthorn, VIC 3122, Australia

*E-mail: dmoss@swin.edu.au

**Abstract**

Realizing optical analogues of quantum phenomena in atomic, molecular, or condensed matter physics has underpinned a range of photonic technologies. Rabi splitting is a quantum phenomenon induced by a strong interaction between two quantum states, and its optical analogues are of fundamental importance for the manipulation of light-matter interactions with wide applications in optoelectronics and nonlinear optics. Here, we propose and theoretically investigate purely optical analogues of Rabi splitting in integrated waveguide-coupled resonators formed by two Sagnac interferometers. By tailoring the coherent mode interference, the spectral response of the devices is engineered to achieve optical analogues of Rabi splitting with anti-crossing behavior in the resonances. Transitions between the Lorentzian, Fano, and Rabi splitting spectral lineshapes are achieved by simply changing the phase shift along the waveguide connecting the two Sagnac interferometers, revealing interesting physical insights about the evolution of different optical analogues of quantum phenomena. The impact of the device structural parameters is also analyzed to facilitate device design and optimization. These results suggest a new way for realizing optical analogues of Rabi splitting based on integrated waveguide-coupled resonators, paving the way for many potential applications that manipulate light-matter interactions in the strong coupling regime.

## I. INTRODUCTION

In atomic, molecular, and condensed matter physics, the response of photon-matter systems is normally characterized via spectroscopic detection of radiation by exploring physical processes like scattering, absorption, or fluorescence, resulting in spectral lineshapes that unveil the nature of the light-matter interactions [1, 2]. This interaction in a confined electromagnetic environment can be controlled to achieve a coupling regime in which coherent energy exchange occurs between light and matter. Such an exchange, in the strong light-matter coupling regime, results in anti-crossing between the atom-like emitter and the cavity-mode dispersion relations, which is described by the so-called Rabi splitting [2-4].

As a physical phenomenon arising from light-matter interactions in the strong coupling regime, Rabi splitting enables quantum coherent oscillations between the joint systems and the quantum superpositions of different quantum states. In the early 1980s, Rabi splitting was observed for many atoms [5]. Since then, it has been widely studied in a variety of quantum and semi-classical systems [2, 3, 5-11], with many applications such as enhancing or modifying the chemical landscape [12, 13], electrical conductivity [14, 15], optical nonlinearity [16], lasing [17, 18], quantum light emission [19, 20], and biological processes [21].

Optical analogues of atomic energy levels allow for the study of complex optical processes using atomic physics concepts, which have resulted in discoveries of photonic crystals (PhCs) [22], topological photonic systems [23, 24], and parity–time symmetric systems [25, 26]. In the past decade, many optical analogues of quantum phenomena have been realized, such as electromagnetically induced transparency [27-29], Fano resonances [30-35], Ramsey interference [36], Mollow triplet [37], and Rabi splitting [38-42]. These optical analogues have been utilized in a variety of applications such as topologically protected lasers [43, 44], light storage [45-49], sensing [50-54], signal multicasting [55, 56], dispersion engineering [57, 58], structured light [59], and photonic computing [60].

In previous work, optical analogues of Rabi splitting have been realized in PhC and plasmonic cavities [6, 39, 61]. In this paper, we propose and theoretically verify a different way for realizing optical analogues of Rabi splitting based on integrated waveguide-coupled resonators. Similar to the manipulation of the interaction between different quantum states in a multi-level atomic system, the coherent mode interference in the waveguide-coupled resonator formed by two Sagnac interferometers is engineered to achieve optical analogues of Rabi splitting with anti-crossing behavior for the resonances. By changing the phase shift along the waveguide connecting the two Sagnac interferometers, transitions from symmetric Lorentzian spectral lineshape to asymmetric Fano and Rabi splitting spectral lineshapes are also achieved, showing interesting trends for the evolutions of different optical analogues of quantum phenomena. Finally, detailed analysis for the impact of device structural parameters is provided to facilitate device design and optimization. Our results theoretically confirm the effectiveness of realizing optical analogues of Rabi splitting based on integrated waveguide-coupled resonators, which offers new possibilities for many potential applications that manipulate light-matter interactions in the strong coupling regime.

## II. DEVICE DESIGN AND OPERATION PRINCIPLE

As schematically illustrated in **Fig. 1(a)**, in the strong light-matter coupling regime where the coherent exchange rate of energy between light and matter is higher than the decay rate, a resonance state $|r>$ (e.g., whispering gallery mode and plasmonic resonances) and a matter excitation state $|e>$ (e.g., excitation states of atoms, molecules, plamons, and quantum dots) strongly couple with each other, resulting in the generation of new hybridized eigenstates separated by the Rabi splitting energy $\hbar\Omega_R$. In contrast to the original independent eigenstates, the new eigenstates arising from field-induced splitting of energy levels show a clear anti-crossing behavior in the spectral response [62], which can be exploited for manipulation of light-matter interaction that has wide applications in low-threshold lasing [63], phase transition modification [39], chemical reactivity tuning [64, 65], Bose-Einstein condensation [66-68], and optical spin switching [69].

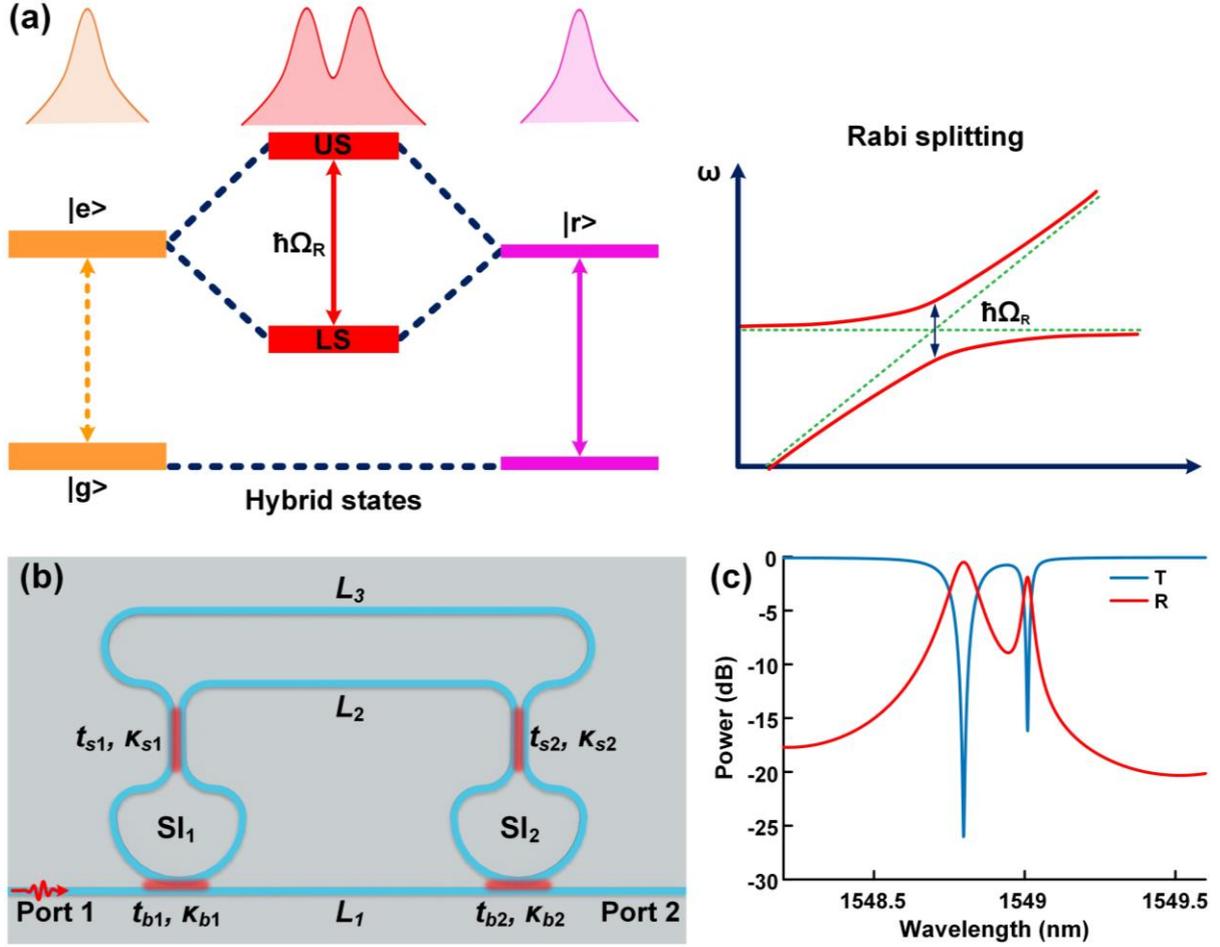

Fig. 1. (a) Schematic illustration of Rabi splitting. |e⟩: matter excitation state. |r⟩: resonance state. |g⟩: ground state. US: upper state. UL: lower state. $\hbar\Omega_R$: Rabi splitting energy, with $\hbar$ denoting the Planck constant and $\Omega_R$ the frequency separation. (b) Schematic configuration of a waveguide-coupled resonator consisting of a bus waveguide side coupled to a resonant loop formed by two Sagnac interferometers. The definitions of $t_{bi}$ ($i$ = 1, 2), $\kappa_{bi}$ ($i$ = 1, 2), $t_{si}$ ($i$ = 1, 2), $\kappa_{si}$ ($i$ = 1, 2), $L_{si}$ ($i$ = 1, 2), and $L_i$ ($i$ = 1–3) are provided in Table I. (c) Power transmission and reflection spectra with input from Port 1 in (b). T: Transmission spectrum at Port 2. R: reflection spectrum at Port 1. The structural parameters are $t_{b1}$ = 0.98, $t_{s1}$ = 0.87, $t_{b2}$ = 0.90, $t_{s2}$ = 0.71, $L_{s1}$ = $L_{s2}$ = 100 µm, and $L_1$ = $L_2$ = $L_3$ =100 µm.

**Fig. 1(b)** shows a schematic of the proposed waveguide-coupled resonator, where a bus waveguide is coupled to a closed resonant loop formed by two Sagnac interferometers. The closed resonant loop couples to the bus waveguide to form two directional couplers, and the interference between the waveguides connecting them is similar to that in a Mach-Zehnder interferometer (which is a finite impulse-response (FIR) filter). On the other hand, the interference between the closed resonant loop and the bus waveguide is similar to that in a ring resonator (which is an infinite impulse-response (IIR) filter). As a result, the device consists of both FIR and IIR filter elements. By introducing Sagnac interferometers in the resonant loop, such a device can also be regarded as a hybrid resonator consisting of both traveling-wave (formed by coherent interference between light waves in the closed-loop resonator and the bus waveguide, similar to ring resonators) and standing-wave (formed by coherent interference induced by reflection between the two Sagnac interferometers, similar to Fabry-Perot cavities) resonator elements. The hybrid nature of the device in **Fig. 1(b)** allows for versatile coherent mode interference that can be tailored for spectral engineering of complex and demanding filtering functions.

The definitions of the device structural parameters are provided in **Table I**. In our following analysis, we investigate the spectral response of the device in **Fig. 1(b)** based on the scattering matrix method [33, 70], using the values of the transverse electric (TE) mode group index $n_g$ = 4.3350 and the propagation loss factor $\alpha$ = 55 m$^{-1}$ (i.e., 2.4 dB/cm) obtained from the fabricated silicon-on-insulator (SOI) devices in our previous work [70, 71]. The devices are designed for, but not limited to, the SOI platform − the principles outlined here are universal for all material platforms.

**Table 1. Definitions of device structural parameters. SI: Sagnac interferometers.**

| Waveguides | Physical length | Transmission factor [a] | Phase shift [b] |
|---|---|---|---|
| Connecting waveguides between $SI_1$ and $SI_2$ ($i$ = 1–3) | $L_i$ | $a_i$ | $\varphi_i$ |
| Sagnac loop in $SI_i$ ($i$ = 1, 2) | $L_{si}$ | $a_{si}$ | $\varphi_{si}$ |
| Directional couplers | | Field transmission coefficient [c] | Field cross-coupling coefficient [c] |
| Coupler in $SI_i$ ($i$ = 1, 2) | | $t_{si}$ | $\kappa_{si}$ |
| Coupler between $SI_i$ and bus waveguide ($i$ = 1, 2) | | $t_{bi}$ | $\kappa_{bi}$ |

[a] $a_i$ = exp(-$\alpha L_i$ / 2), $a_{si}$ = exp(-$\alpha L_{si}$ / 2), where $\alpha$ is the power propagation loss factor.
[b] $\varphi_i$ = $2\pi n_g L_i$ / $\lambda$, $\varphi_{si}$ = $2\pi n_g L_{si}$ / $\lambda$, where $n_g$ is the group index and $\lambda$ is the wavelength.
[c] $t_{si}^2 + \kappa_{si}^2$ = 1 and $t_{bi}^2 + \kappa_{bi}^2$ = 1 for lossless coupling is assumed for all the directional couplers.

We tailor the spectral response of the device in **Fig. 1(b)** with bidirectional light propagation to realize optical analogues of Rabi splitting. The power transmission and reflection spectra with input from Port 1 are depicted in **Fig. 1(c)**. Unless elsewhere specified, the spectral response of the device is assumed to have an input from Port 1 in **Fig. 1(b)**, with the transmitted light at Port 2 and the reflected light back to Port 1. In **Fig. 1(c)**, the device structural parameters are $t_{b1} = 0.98$, $t_{s1} = 0.87$, $t_{b2} = 0.90$, $t_{s2} = 0.71$, $L_{s1} = L_{s2} = 100$ μm, and $L_1 = L_2 = L_3 = 100$ μm. As can be seen, both the transmission spectra at Port 2 and the reflection spectra at Port 1 show Rabi-like splitting that features asymmetric split resonances, which result from coherent mode interference, analogous with the interaction between different quantum states in multi-level atomic systems. Similar Rabi splitting spectra in the visible and mid-infrared regions have also been observed for PhC and plasmonic nanocavities [6, 39]. However, compared to these, the fabrication of waveguide-coupled resonators does not require high lithography resolution and shows higher tolerance to fabrication imperfections such as lithographic smoothing effects and quantization errors due to the finite grid size [70], which makes it much easier to accurately engineer the spectral response and overall device performance.

## III. ANTI-CROSSING OF SPLIT RESONANCES

In this section, we investigate the resonance anti-crossing behavior of the device in **Fig. 1(b)**, which is a key feature of Rabi splitting. A typical application of the anti-crossing behavior of the split resonances is to engineer the dispersion of the devices to generate artificial anomalous dispersion in devices that exhibit intrinsic normal dispersion [57, 58, 72], for applications to nonlinear optics. We also provide an analysis for the influence of the device structural parameters on the Rabi split resonances. To simplify the comparison, we vary only one structural parameter in each figure of this section, while keeping the others the same as in **Fig. 1(c)**.

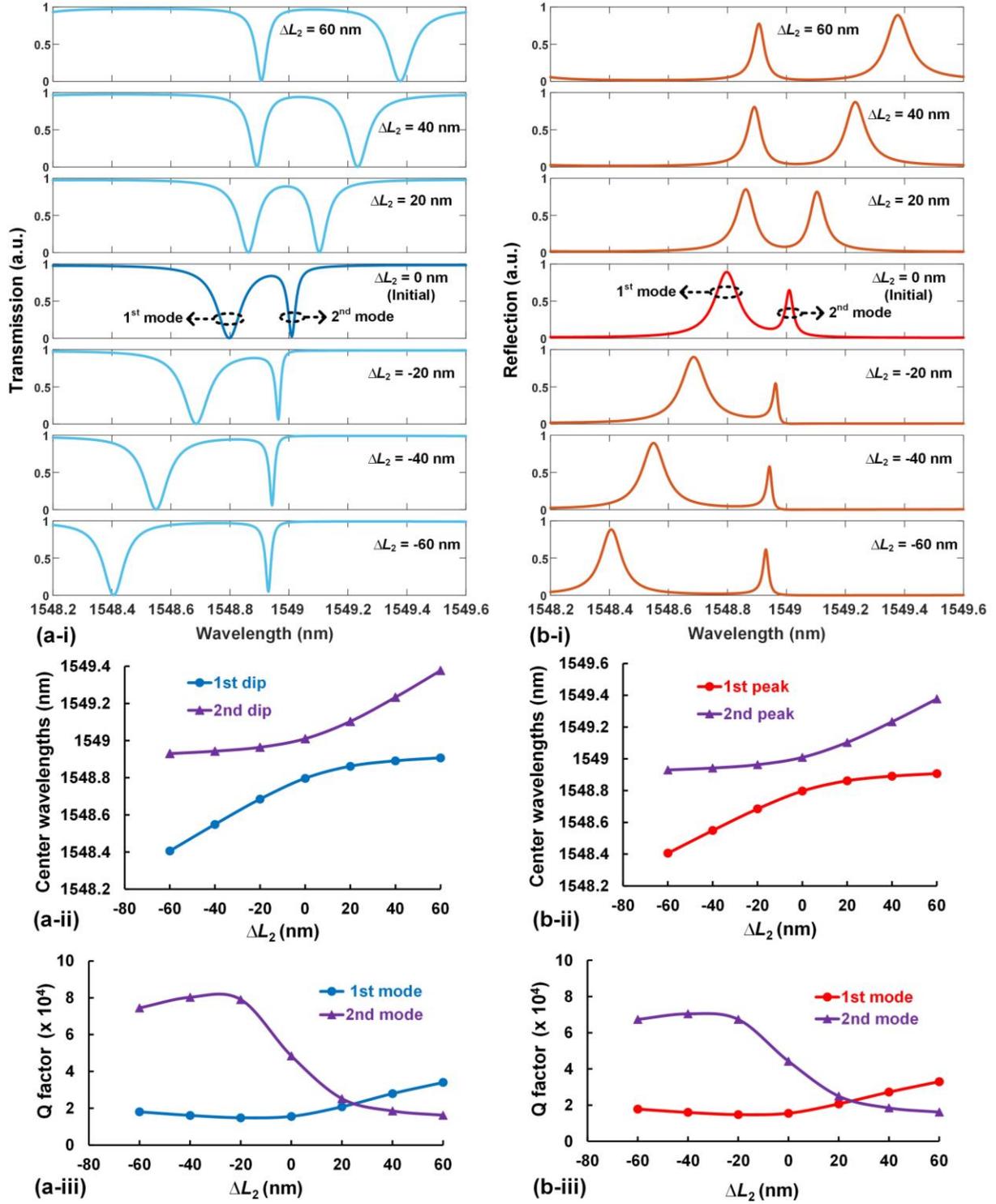

Fig. 2. Influence of the variation in the length of $L_2$ ($\Delta L_2$) on the device spectral response. (a) Rabi split resonances for various $\Delta L_2$ at the transmission port. (b) Rabi split resonances for various $\Delta L_2$ at the reflection port. In (a) and (b), (i) shows the power transmission spectra, (ii) shows the center wavelengths of the two resonances in (i) as functions of $\Delta L_2$, and (iii) shows the Q factors of the two resonances in (i) as functions of $\Delta L_2$. The structural parameters are kept the same as those in Fig. 1(c) except for the variation in the length of $L_2$.

In **Fig. 2** we compare the device spectral response for various $\Delta L_2$, i.e., the variation in the length of $L_2$. As $\Delta L_2$ changes, both the transmission spectra in **Fig. 2(a-i)** and the reflection spectra in **Fig. 2(b-i)** show obvious anti-crossing behavior in the split resonances when the two resonance modes approach each other. The center wavelengths and Q factors of the two resonance modes in **Fig. 2(a-i)** as functions of $\Delta L_2$ are depicted in **Figs. 2(a-ii)** and 2**(a-iii)**, respectively. The corresponding results for the two resonance modes in **Fig. 2(b-i)** are shown in **Figs. 2(b-ii)** and 2**(b-iii)**. As can be seen, both the center wavelengths and the Q factors of the two split resonances show obvious changes with $\Delta L_2$. This reflects the strong interaction between the two modes resulting from coherent mode interference, analogous to the coupling between the resonance and excitation states in **Fig. 1(a)** in the strong light-matter coupling regime.

**Fig. 3** shows the device spectral response for various $\Delta L_3$, i.e., the variation in the length of $L_3$. Similar to **Fig. 2**, the two Rabi split resonances avoid crossing each other in both the transmission (**Fig. 3(a-i)**) and reflection spectra (**Fig. 3(b-i)**). In both, the Q factors of the first resonance decreases with $\Delta L_3$, while that of the second resonance increases with $\Delta L_3$, showing the opposite trend to **Figs. 2(a-iii)** and **(b-iii)**.

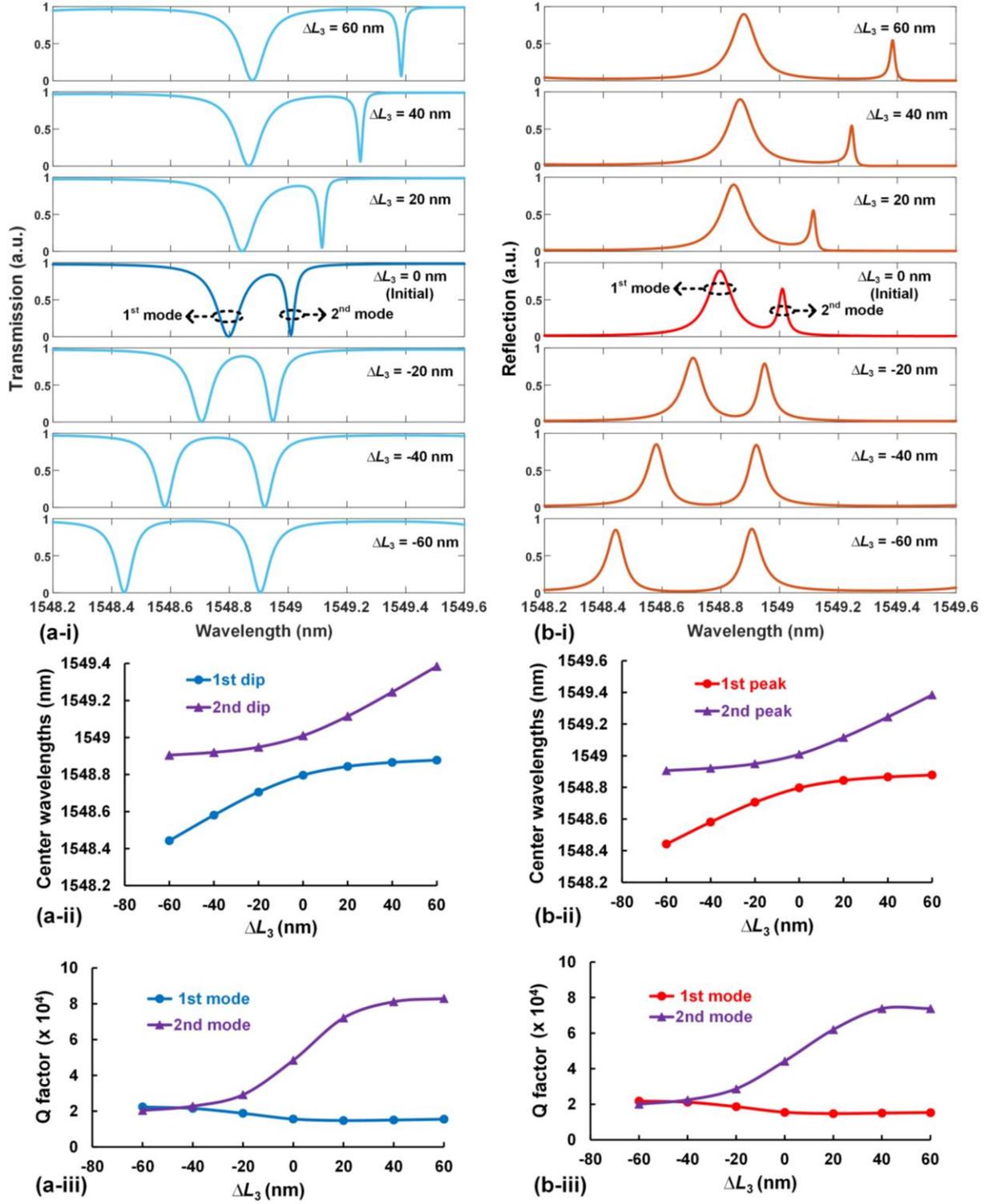

Fig. 3. Influence of the variation in the length of $L_3$ ($\Delta L_3$) on the device spectral response. (a) Rabi split resonances for various $\Delta L_3$ at the transmission port. (b) Rabi split resonances for various $\Delta L_3$ at the reflection port. In (a) and (b), (i) shows the power transmission spectra, (ii) shows the center wavelengths of the two resonances in (i) as functions of $\Delta L_3$, and (iii) shows the Q factors of the two resonances in (i) as functions of $\Delta L_3$. The structural parameters are kept the same as those in Fig. 1(c) except for the variation in the length of $L_3$.

In **Fig. 4**, we compare the device spectral response for various $\Delta L_1$, which is the length variation in $L_1$. In both the transmission and reflection spectra, the center wavelengths of the two split resonances remain almost unchanged, mainly because $L_1$ corresponds to the waveguide outside the closed resonant loop. In contrast, the Q factors of the two Rabi split resonances show obvious differences, with the Q factors of the first and second resonances decreasing and increasing with $\Delta L_1$, respectively. This is similar to the trend for the changes of the Q factors with $\Delta L_3$ in **Figs. 3(a-iii)** and **(b-iii)**.

In practical applications, different values of $\Delta L_i$ ($i$ = 1, 2, 3) in **Figs. 2 – 4** can be achieved by varying the physical lengths of the corresponding waveguides in passive devices. By introducing thermo-optic micro-heaters [32, 60] or PN junctions [73, 74] along the corresponding waveguides to tune their phase shifts, the changes of $\Delta L_i$ ($i$ = 1, 2, 3) can also be realized via active tuning of passive devices, which allows for tunable Rabi splitting characteristics to meet the requirements of different applications.

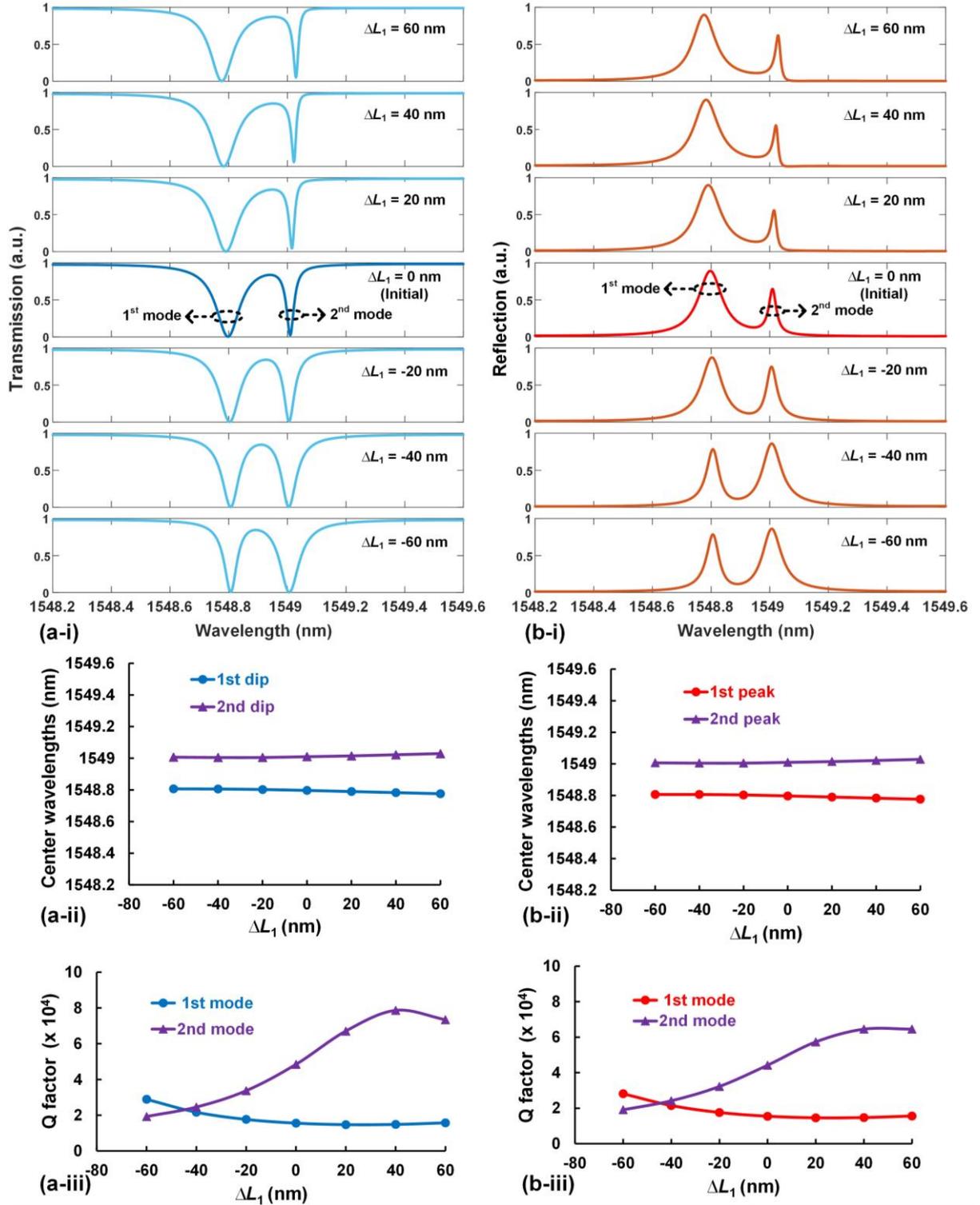

Fig. 4. Influence of the variation in the length of $L_1$ ($\Delta L_1$) on the device spectral response. (a) Rabi split resonances for various $\Delta L_1$ at the transmission port. (b) Rabi split resonances for various $\Delta L_1$ at the reflection port. In (a) and (b), (i) shows the power transmission spectra, (ii) shows the center wavelengths of the two resonances in (i) as functions of $\Delta L_1$, and (iii) shows the Q factors of the two resonances in (i) as functions of $\Delta L_1$. The structural parameters are kept the same as those in Fig. 1(c) except for the variation in the length of $L_1$.

In addition to phase shifts along the connecting waveguides, we investigate the influence of the coupling strengths of the directional couplers on the Rabi split resonances in **Figs. 5** and **6**. Since their influence on the center wavelengths of the split resonances is not as significant as varying $\Delta L_i$ ($i$ = 1, 2, 3) in **Figs. 2 – 4**, we plot the spectra for different coupling strengths in the same figure to highlight the differences.

We first investigate the influence of the coupling strengths of the directional couplers in the two Sagnac interferometers (i.e., $t_{s1}$ and $t_{s2}$), which determines the reflectivity of the Sagnac interferometers. **Fig. 5(a)** shows the transmission and reflection spectra for various $t_{s1}$. As can be seen, the extinction ratios of the Rabi split resonances (defined as the difference between the maximum and minimum transmission) remain almost unchanged, while the spectral interval between the split resonances increases with $t_{s1}$. Similar trends are also observed in **Fig. 5(b)** that plot the transmission and reflection spectra versus $t_{s2}$. These result from the fact that changes in $t_{s1}$ or $t_{s2}$ alter the mutual coupling between the light waves propagating in two opposite directions, which varies the degree of Rabi mode splitting that have various spectral intervals between the split resonances.

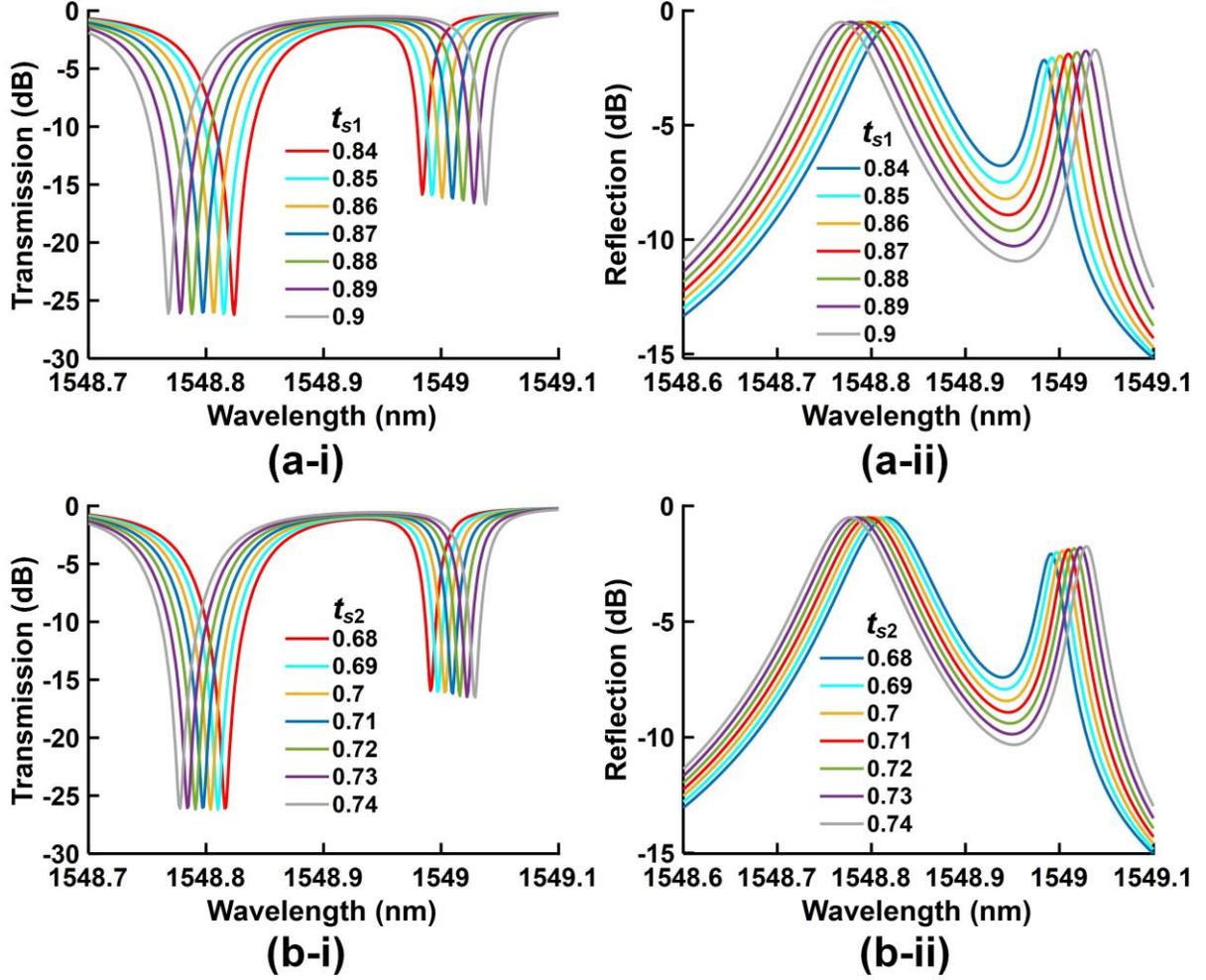

Fig. 5. (a) Influence of $t_{s1}$ on Rabi split resonances at (i) the transmission port and (ii) the reflection port. (b) Influence of $t_{s2}$ on Rabi split resonances at (i) the transmission port and (ii) the reflection port. In (a) and (b), the structural parameters are kept the same as those in Fig. 1(c) except for $t_{s1}$ and $t_{s2}$, respectively.

In **Fig. 6**, we investigate the influence of varying the strengths of the directional couplers between the bus waveguide and Sagnac interferometers (i.e., $t_{b1}$ and $t_{b2}$), i.e., the energy coupling strength between the closed resonant loop and bus waveguide. **Fig. 6(a)** shows the transmission and reflection spectra versus $t_{b1}$, where the left resonance extinction ratio decreases with $t_{b1}$, while the right extinction ratio shows the opposite trend, reflecting their trade-off. The transmission and reflection spectra versus $t_{b2}$ are shown in **Fig. 6(b)**. Unlike the trade-off between the extinction ratios of the two resonances in **Fig. 6(a)**, both resonance extinction ratios in **Fig. 6(b)** decrease with $t_{b2}$, highlighting the difference in the influence of $t_{b1}$ and $t_{b2}$ on the transmission and reflection spectra. Although $t_{b1}$ and $t_{b2}$ have more influence on the extinction ratios of the Rabi split resonances than $t_{s1}$ and $t_{s2}$, their influence on the spectral interval between the split resonances is not as significant. This is mainly because the spectral interval between the split resonances, which reflects the degree of Rabi mode splitting, is determined by the coupling strength between the bidirectional light waves within the closed resonant loop. On the other hand, the extinction ratios of the split resonances are determined by the energy exchange between the resonant loop and the bus waveguide, which is similar to the different coupling regimes in ring resonators [75].

In practical applications, dynamic tuning of the coupling strength of the directional couplers in **Figs. 5** and **6** can be realized by using Mach-Zehnder interferometric couplers to replace the directional couplers and adjusting the phase difference between the two arms. Recently, compact tunable directional couplers have also been demonstrated by directly integrating thermo-optic micro-heaters above the coupling regions [76, 77], which induces thermal gradients that lead to phase velocity mismatch between the coupled modes of the waveguides, enabling dynamic tuning of the coupling strength.

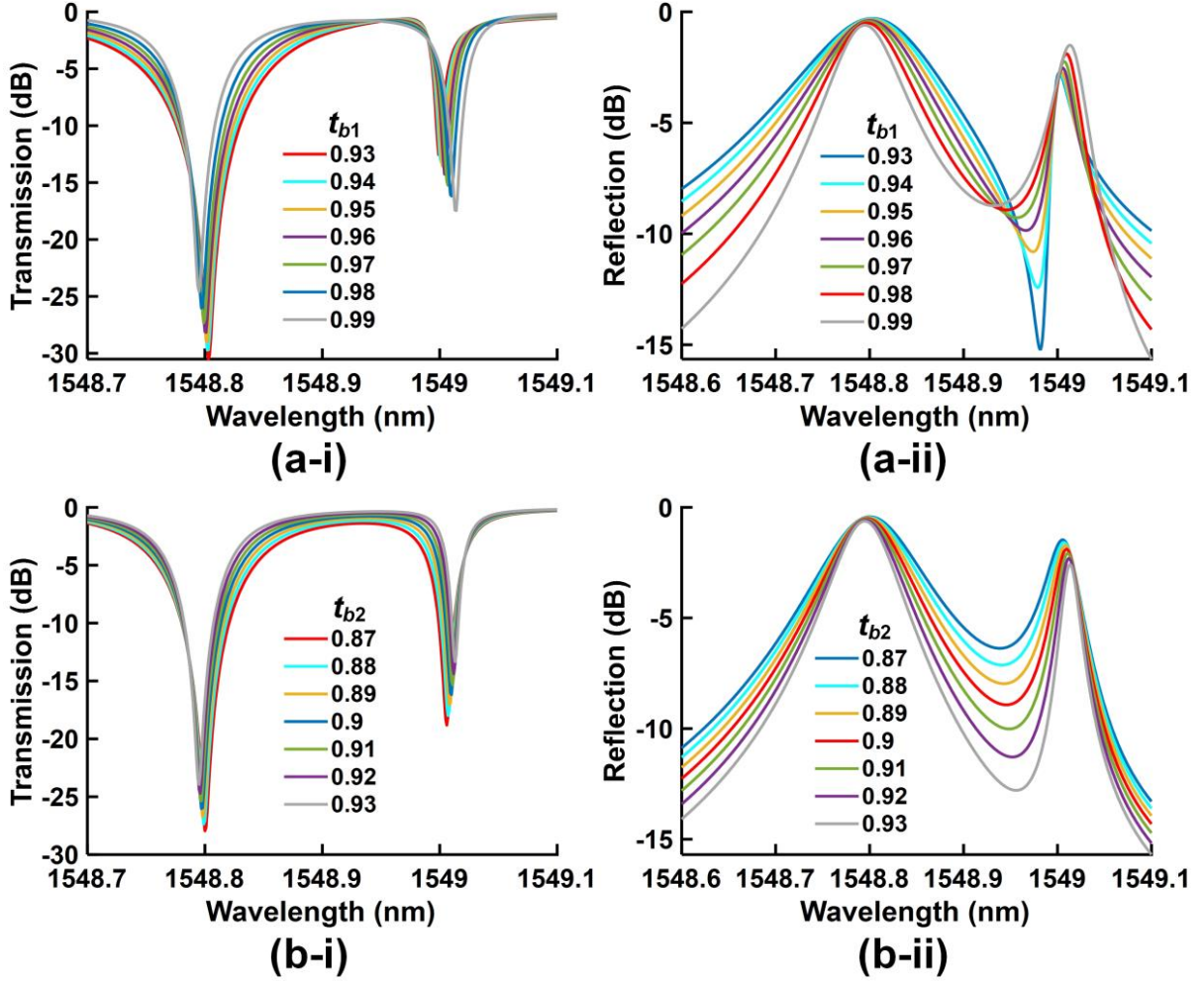

Fig. 6. (a) Influence of $t_{b1}$ on Rabi split resonances at (i) the transmission port and (ii) the reflection port. (b) Influence of $t_{b2}$ on Rabi split resonances at (i) the transmission port and (ii) the reflection port. In (a) and (b), the structural parameters are kept the same as those in Fig. 1(c) except for $t_{b1}$ and $t_{b2}$, respectively.

## IV. TRANSITIONS BETWEEN LORENTZIAN, FANO, AND RABI SPLITTING SPECTRAL LINESHAPES

The spectroscopic detection of radiation via physical processes such as scattering, absorption, or fluorescence has been widely used for characterizing the response of photon-matter systems, resulting in featured spectral lineshapes that unveil the fundamentals of light-matter interactions [1, 2]. In addition to Rabi splitting, there are other typical resonance spectral lineshapes. For example, the symmetric Lorentzian spectral lineshape, which describes the finite radiative lifetimes of excited states, and is commonly seen in the response spectra of ring resonators and Fabry-Perot cavities [71, 73, 78-80]. In addition, Fano resonances, which feature an asymmetric spectral lineshape induced by interference between a discrete quantum state and a continuum band of states [81, 82], have underpinned many applications such as switching [83], sensing [84, 85], lasing [86], and directional scattering [87, 88]. Despite originating from different underlying physics, the different spectral lineshapes are related to each other and sometimes can transit from one to another. In this section, we engineer the spectral response of the device in **Fig. 1(b)** to achieve transitions between Lorentzian, Fano, and Rabi splitting resonance spectral lineshapes.

**Fig. 7** shows the power transmission and reflection spectra for different values of $L_3$. The device structural parameters are $L_{s1} = L_{s2} = 100$ μm, $L_1 = 200$ μm, $L_2 = 100$ μm, $t_{b1} = 0.83$, $t_{s1} = 0.707$, $t_{b2} = 0.98$, and $t_{s2} = 0.707$. In **Fig. 7**, there is a single resonance with a symmetric Lorentzian spectral lineshape in both the transmission and reflection spectra when $L_3 = 100$ μm. As $L_3$ increases, the single resonance gradually splits to form a Fano asymmetric spectral lineshape in the reflection spectrum (when $L_3 = 100.02$ μm), and then evolves to become the Rabi splitting lineshape (when $L_3 = 100.03$ μm and $100.04$ μm). By further increasing $L_3$, the split resonances gradually evolve to form a single symmetric Lorentzian spectral lineshape (when $L_3 = 100.17$ μm). This illustrates how transitions between the Lorentzian, Fano, and Rabi splitting resonance lineshapes can be realized by simply adjusting the phase shift along $L_3$. Another interesting phenomenon is that both Fano and Rabi splitting spectral lineshapes can be obtained for the same device with $L_3 = 100.02$ μm, where the Fano and Rabi splitting spectral lineshapes are achieved for the resonances observed in the reflection and transmission directions, respectively. Similarly, the transitions among the Lorentzian, Fano, and Rabi splitting resonance lineshapes can also be realized by changing $L_2$, as shown in **Fig. 8**.

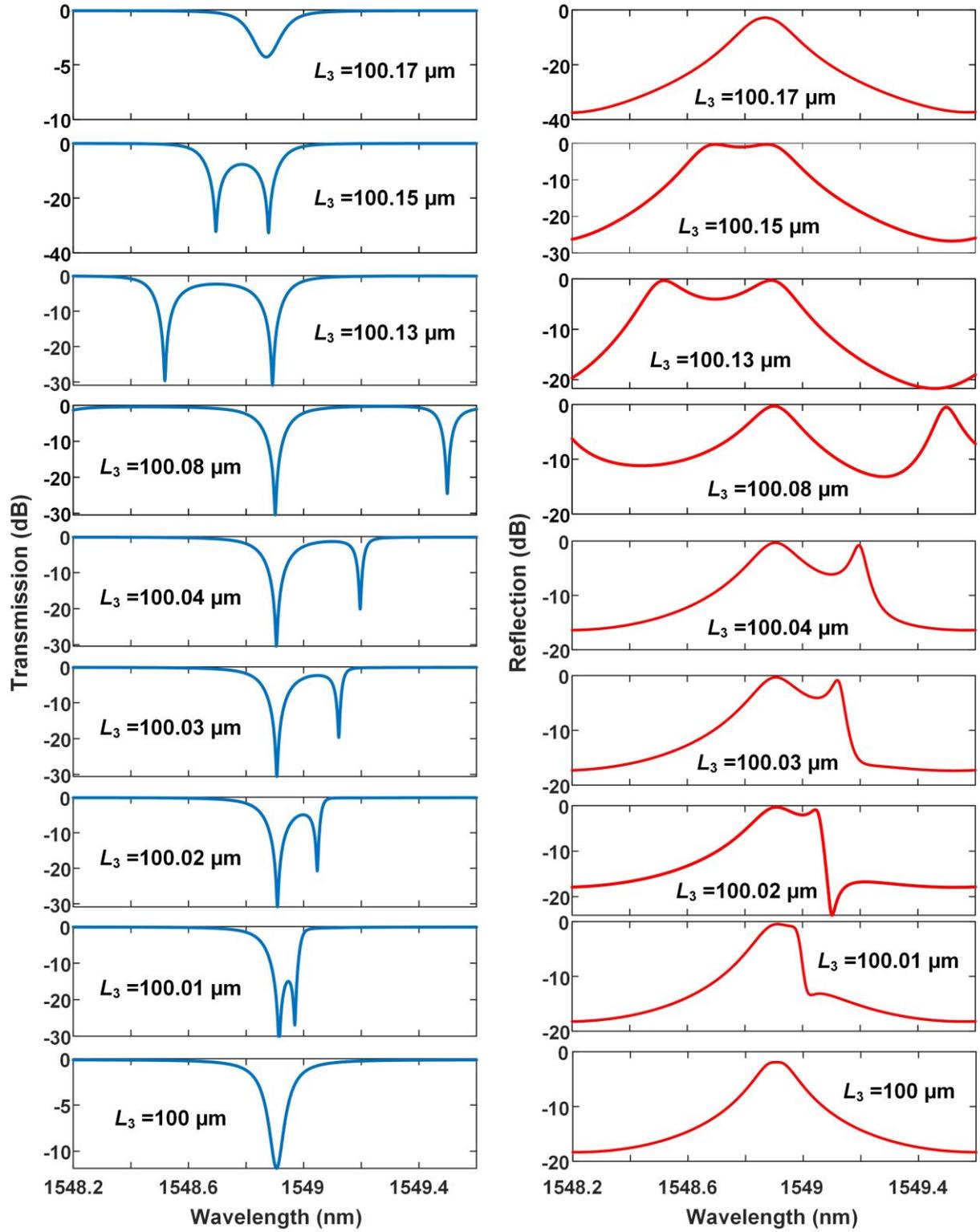

Fig. 7. Power transmission (left) and reflection (right) spectra for various $L_3$. The structural parameters are $t_{b1} = 0.83$, $t_{s1} = 0.707$, $t_{b2} = 0.98$, $t_{s2} = 0.707$, $L_{s1} = L_{s2} = 100$ μm, $L_1 = 200$ μm, and $L_2 = 100$ μm.

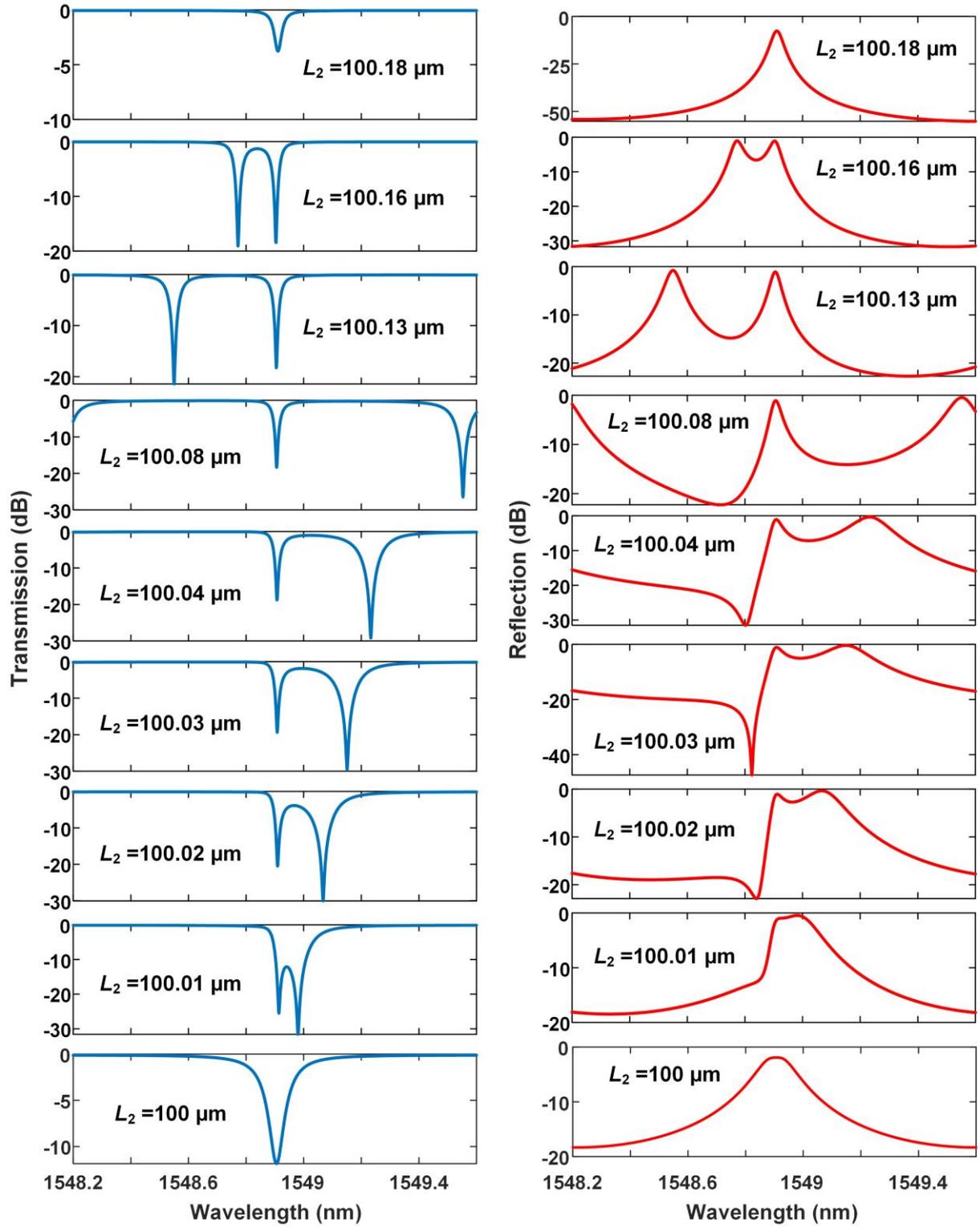

Fig. 8. Power transmission (left) and reflection (right) spectra for various $L_2$. The structural parameters are $t_{b1} = 0.83$, $t_{s1} = 0.707$, $t_{b2} = 0.98$, $t_{s2} = 0.707$, $L_{s1} = L_{s2} = 100$ μm, $L_1 = 200$ μm, and $L_3 = 100$ μm.

The versatility and degree of control achieved using these advanced design techniques based on Sagnac interferometers [89-91] will likely have wide applications to many different areas beyond linear optical filters, including nonlinear optics and microcomb based devices [92-148] as well as quantum optical photonic chips [149-161] and photonic integrated chips based on novel 2D materials and structures [162-184].

## V. CONCLUSIONS

We theoretically investigate optical analogues of Rabi splitting in integrated waveguide-coupled resonators formed by two Sagnac interferometers. Coherent mode interference in the proposed device is tailored to achieve optical analogues of Rabi splitting with a clear avoided behavior for the resonances. Moreover, transitions between the Lorentzian, Fano, and Rabi splitting resonance lineshapes are achieved by changing the phase shift along the connecting waveguide between the two Sagnac interferometers. A detailed analysis for the influence of the structural parameters on the device performance is also provided. Our work offers new possibilities for realizing optical analogues of Rabi splitting based on integrated waveguide-coupled resonators and provides new prospects for realizing optical analogues of quantum physics by exploiting Sagnac interference in integrated photonic devices. This will allow for attractive benefits in providing devices with compact footprint, high scalability, high fabrication tolerance, and mass producibility for practical applications.

**Competing Interests**

The authors declare that there are no competing interests.